\documentclass[pdflatex,sn-mathphys-num]{sn-jnl}

\usepackage{graphicx}%
\usepackage{multirow}%
\usepackage{amsmath,amssymb,amsfonts}%
\usepackage{amsthm}%
\usepackage{mathrsfs}%
\usepackage[title]{appendix}%
\usepackage{xcolor}%
\usepackage{textcomp}%
\usepackage{manyfoot}%
\usepackage{booktabs}%
\usepackage{algorithm}%
\usepackage{algorithmicx}%
\usepackage{algpseudocode}%
\usepackage{listings}%

\theoremstyle{thmstyleone}%
\theoremstyle{thmstyletwo}%

\theoremstyle{thmstylethree}%

\begin{document}

\title[Article Title]{Integrated thin film lithium niobate mid-infrared modulator}


\author*[1]{\fnm{Pierre} \sur{Didier}}\email{pdidier@phys.ethz.ch}

\author[1]{\fnm{Prakhar} \sur{Jain}}

\author[2]{\fnm{Mathieu} \sur{Bertrand}}

\author[1]{\fnm{Jost} \sur{Kellner}}

\author[1]{\fnm{Oliver} \sur{Pitz}}

\author[3]{\fnm{Zhecheng} \sur{Dai}}

\author[2]{\fnm{Mattias} \sur{Beck}}

\author[3]{\fnm{Baile} \sur{Chen}}

\author[2]{\fnm{Jérome} \sur{Faist}}

\author[1]{\fnm{Rachel} \sur{Grange}}

\affil[1]{ETH Zurich, Department of Physics, Institute for Quantum Electronics, Optical Nanomaterial Group, Zurich, Switzerland}
\affil[2]{ETH Zurich, Department of Physics, Institute for Quantum Electronics, Quantum Optoelectronics Group, Zurich, Switzerland}
\affil[3]{School of Information Science and Technology, ShanghaiTech University, Shanghai 201210, China}

\keywords{integrated photonics, lithium niobate, mid-infrared, amplitude and phase optical modulation}


\abstract{The mid-infrared spectral range holds great promise for applications such as molecular spectroscopy and telecommunications. Many key molecules exhibit strong absorption features in this range, and free-space optical communication benefits from reduced atmospheric attenuation and low transmission losses in specific wavelength bands spanning from 3 to 14\,\textmu m. Recent progress in MIR photonics has been fuelled by the rapid development of efficient light sources and detectors. However, further advancement is hindered by the lack of low-loss, high-performance integrated photonic platforms and modulators. Lithium niobate on sapphire is a promising candidate, operating across a broad spectral range from 0.4\,\textmu m to 4.5\,\textmu m. We demonstrate a broadband, high-speed lithium niobate on sapphire Mach--Zehnder electro-optic modulator operating from 3.95 to 4.3\,\textmu m. The device achieves a 3\,dB bandwidth exceeding 20\,GHz, an extinction ratio of 34\,dB, and a half-wave voltage of 22\,V$\cdot$cm, delivering optical output power at the half-milliwatt level. These properties are leveraged to demonstrate data transmission at 10\,Gbit/s. The modulator is also used to generate a frequency comb with a width of 80 GHz. Furthermore, we demonstrate full $\pi$-phase modulation in the MIR, representing a key milestone for integrated MIR photonics. These results establish a pathway toward high-speed, energy-efficient MIR photonic systems for applications in telecommunications, sensing, and quantum technologies.}

\maketitle
\section{Main}\label{sec1}

Mid-infrared (MIR) photonics require the development of highly efficient platforms to fully exploit the unique advantages of this spectral region. The MIR spectrum, spanning 3-14\,\textmu m, offers significant benefits for free-space optical applications~\cite{Jony2019,Spitz2021}, including high atmospheric transparency, reduced scattering from micrometric aerosols, and enhanced resilience to atmospheric turbulence \cite{corrigan2009quantum,liu2019mid,Esmail2017}. Furthermore, many environmentally relevant molecules exhibit strong absorption features in the MIR that are up to an order of magnitude stronger than in the near-infrared, enabling more precise and sensitive detection \cite{haas2016advances,zhang2014applications}. In this context, phase and amplitude modulation are essential techniques for MIR spectroscopy - improving sensitivity, resolution, and signal-to-noise ratio.~\cite{schilt2003wavelength} There have been many efforts towards amplitude modulation in the MIR, each with their own drawbacks. The most common demonstrations have been with direct modulation of quantum or interband cascade lasers (QCLs/ICLs)\cite{Faist1994,Meyer1995,yang1997high}. QCLs operate most reliably in the long-wave infrared (3.8--14\,\textmu m) range. While RF packaging requirements can introduce thermal constraints, QCLs have been directly modulated at high speeds. Optical 3\,dB bandwidths up to 3\,GHz have been demonstrated~\cite{dely2024unipolar,hinkov2016rf}, but the modulation depth remains relatively low (approximately 10\,dB) due to strong nonlinearities at high drive currents, along with the need for large current swings and high electrical power. ICL-based schemes have demonstrated gigahertz-scale performance~\cite{didier2023interband}, benefiting from less demanding thermal management requirements, but they suffer from limited output power. Moreover, direct modulation of QCLs/ICLs inherently causes simultaneous amplitude and phase modulation, limiting their use in many applications. Alternative approaches such as external free space Stark effect modulators have achieved comparable bandwidths~\cite{bonazzi2024metamaterial,pirotta2021fast,dely2022} and extinction ratios up to 3\,dB~\cite{didier2022high}, yet also face scalability limitations due to highly sensitive free space coupling. Ultimately, a weak electro-optic effect and a lack of MIR-compatible materials have caused integrated MIR modulators in the 3--5\,\textmu m range to remain limited, with bandwidths in the sub-gigahertz range~\cite{li2019ge}, very high losses~\cite{nedeljkovic2019silicon}, and shallow modulation depths~\cite{montesinos2022mid}.
Phase modulation in the mid-infrared thus far has been demonstrated in a few platforms, including optical parametric oscillators (OPOs) based on periodically-poled bulk lithium niobate~\cite{zou2022high,zhou2024demonstration}, Stark effect-based devices using intersubband transitions~\cite{dely2023heterodyne,chung2023electrical} or Graphene-based structures~\cite{sherrott2017experimental}. While OPOs provide high optical power, they rely on bulky tabletop components and have very high power consumption. Other approaches have not been able to simultaneously achieve a full $\pi$ phase shift, and a high extinction ratio (typically around $\sim$1\,dB), whereby their requirement of free-space configurations also poses significant challenges for scalability.
Lithium niobate (LiNbO\textsubscript{3} or LN) presents a promising alternative for the 3-5 \textmu m wavelength range, offering low propagation loss, a wide transparency window (extending beyond 4.3\,\textmu m), and a strong second-order non-linearity $\mathrm{\chi^{(2)}}$, with a large Pockels coefficient up to 33 pm/V~\cite{boyd2008nonlinear}. This enables a change of the refractive index via an externally applied electric field. Conventional bulk LN waveguides are based on titanium in-diffusion and suffer from weak mode confinement and large bend radii, which limit both nonlinear efficiency and potential for scalable integration~\cite{xie2022linbo3}. The advent of the thin film lithium niobate on insulator (LNOI) platform has enabled high-confinement waveguides and compact, high-performance modulators in the C-band reaching bandwidths above 100 GHz with $\mathrm{V}_{\pi} \mathrm{L}$ below 1\,V~\cite{wang2018integrated,xu2020high}. However, the LNOI material stack includes a buried silicon dioxide (SiO$_2$) layer, which absorbs strongly beyond 3.4\,\textmu m~\cite{soref2006silicon}, hence limiting MIR operation. This limitation can be addressed by the lithium niobate on sapphire (LNOS) platform~\cite{Mishra2021}, which replaces the SiO\textsubscript{2} substrate with sapphire, thereby extending the transparency window up to 4.5\,\textmu m. LNOS preserves the electro-optic advantages of lithium niobate while enabling low-loss, high-confinement waveguides in the MIR. Moreover, sapphire offers several attractive properties, including strong acousto-optic effects, low radio frequency loss and excellent thermal management~\cite{dobrovinskaya2009properties}.

\begin{figure}[ht!]
    \centering
    \includegraphics[width=13cm]{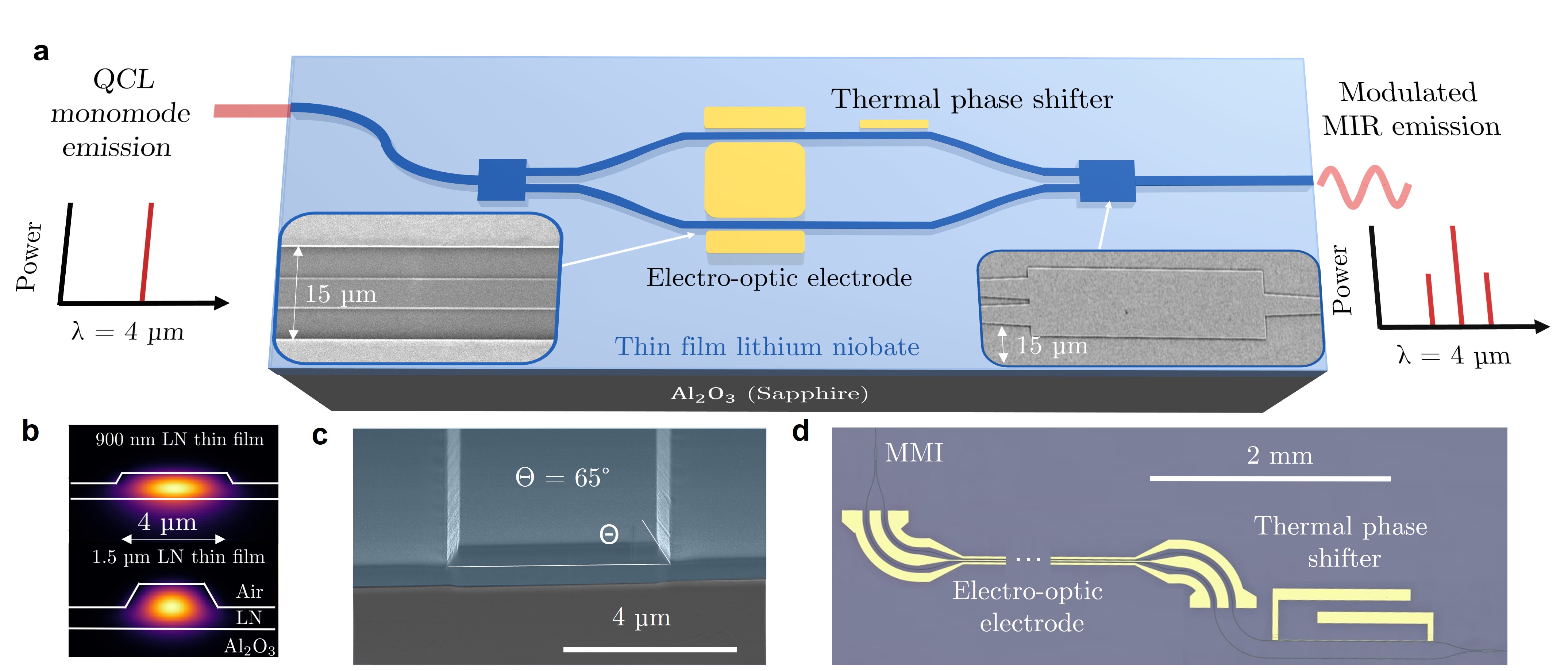}
    \caption{\textbf{LNOS Mach–Zehnder modulator.} \textbf{a,} Schematic of the Mach–Zehnder optical amplitude modulator (MZM). The insets show SEM images of the multi-mode interferometer (bottom right) and electrodes around the waveguide (bottom left). \textbf{b,} Cross-sectional simulations showing optical mode confinement within lithium niobate thin films of thickness 0.9\,\textmu m (top) and 1.5\,\textmu m (bottom) at a wavelength of 4\,\textmu m. The 0.9\,\textmu m film supports only the fundamental transverse electric (TE) mode, while the 1.5\,\textmu m film supports both a TE and a transverse magnetic (TM) mode. \textbf{c,} SEM image of a waveguide facet, mechanically polished and subsequently refined using focused ion beam (FIB) milling. \textbf{d,} Optical microscope image of a fabricated device, showing the MMI coupler (top left), electro-optic electrodes (middle), and thermal phase shifter (right).} 
    \label{fig:1}
\end{figure}

We demonstrate an integrated amplitude modulator in the MIR, operating near a wavelength of 4~\textmu m. The modulators presented in this work are based on a travelling-wave Mach-Zehnder interferometer architecture implemented on the LNOS platform. Light is split by a multi-mode interferometer (MMI) into two parallel waveguide arms positioned within the gap of a ground–signal–ground coplanar microwave line, where tightly confined optical and radio-frequency fields propagate in the same direction. The differential electric field across the arms induces an antisymmetric phase shift via the Pockels effect, enabling amplitude modulation through interference at the output MMI junction. Our modulator exhibits a half-wave voltage-length product $\mathrm{V}_{\pi} \mathrm{L}$ of 22\,V.cm, while achieving a 3-dB bandwidth of higher than 20 GHz with an extinction ratio of 34 dB and an insertion loss of 14.1~dB. To our knowledge, this represents the first photonic integrated MZM demonstrated in the MIR, surpassing previous demonstrated modulators in terms of speed and extinction ratio by nearly two orders of magnitude~\cite{XuDongZhongZhengQiuZhaoJiaLeeHu2023}. We also compare the performance of two device geometries based on different LN film thicknesses, namely 0.9\,\textmu m and 1.5\,\textmu m. Crucially, this work demonstrates the first integrated high-speed $\mathrm{\pi}$-phase shift modulation in the mid-infrared. The device exhibits strong modulation depth and excellent signal integrity at high frequencies, with an output optical power exceeding half a milliwatt. Finally, we demonstrate that the modulator maintains effective operation up to a wavelength of 4.3\,\textmu m without considerable degradation in performance. These results represent a significant advancement in integrated mid-infrared photonics, enabling new capabilities in coherent sensing, high-resolution spectroscopy, and free-space optical communication.

\section{Result}\label{sec_Result}

\begin{figure}[ht!]
    \centering
    \includegraphics[width=13cm]{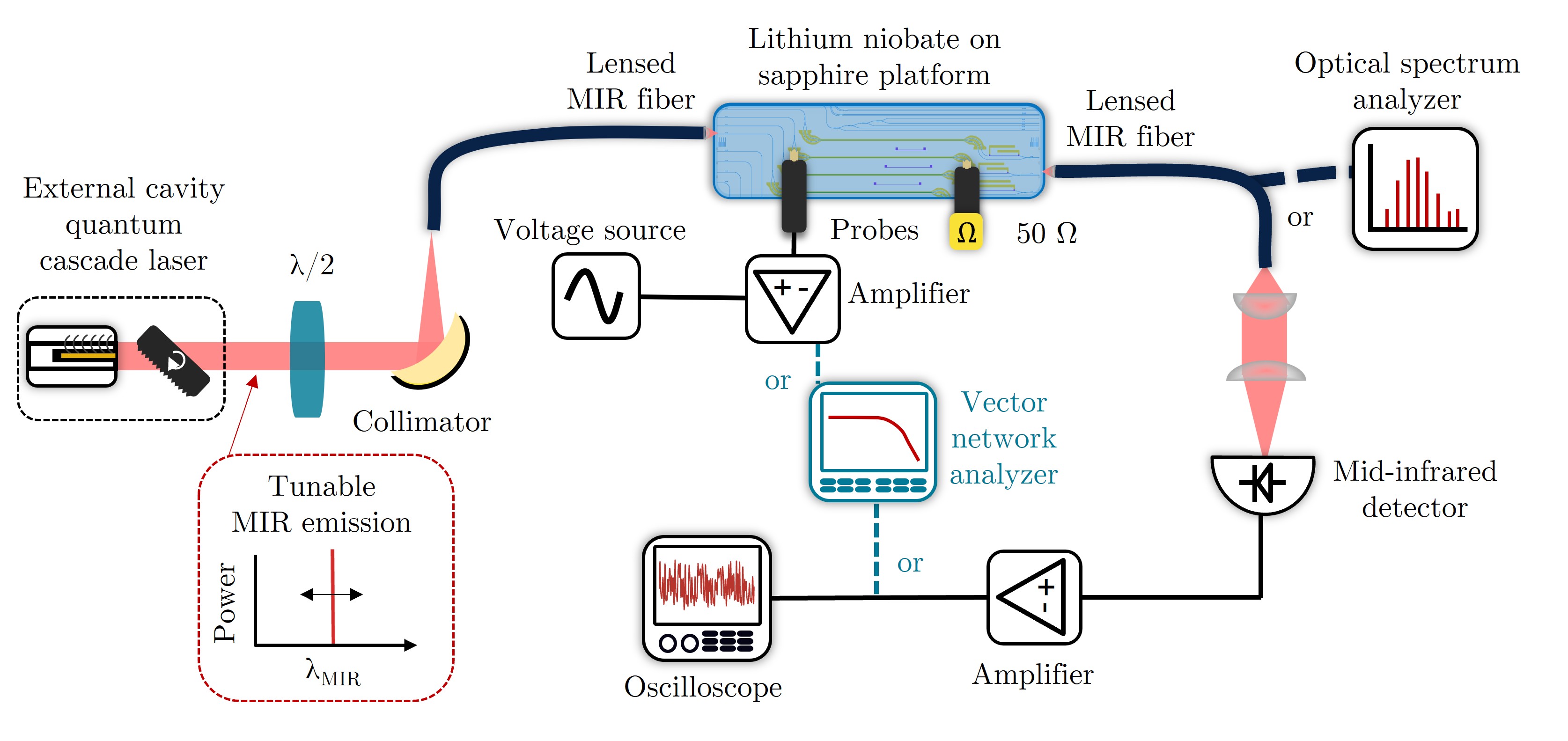}
    \caption{\textbf{Schematic of the characterization setup for the MIR integrated MZM.} The output of an external cavity laser around 4\,\textmu m is coupled into a MIR lensed fiber and then into the photonic chip. At the output, the light is collected using another lensed fiber. The output from the fiber is either directly injected into an optical spectrum analyzer (OSA), a power meter, or collimated and focused onto a high-speed MIR photodetector. This setup can be adapted for both static and dynamic characterization.} 
    \label{fig:2}
\end{figure}

\subsection{Fabrication and design of the lithium niobate on sapphire platform}

A schematic of the device layout is shown in Fig.~\ref{fig:1}a. The chips were fabricated from commercially available 0.9-\textmu m-thick and 1.5-\textmu m x-cut LNOS wafers. Photonic waveguides were patterned using electron-beam lithography and etched via argon-based reactive ion etching. The fabricated waveguides feature a top width of 4\,\textmu m with an etch depth of 400\,nm for the 0.9\,\textmu m-thick film, and a top width of 2.5\,\textmu m with an etch depth of 920\,nm for the 1.5\,\textmu m-thick film. In both cases, the sidewall angle is approximately 65$^\circ$, resulting from the anisotropic physical etching process~\cite{kaufmann2023redeposition}. These designs ensure single-mode transverse electrical operation at a wavelength of 4~\textmu m, corresponding to the target operating range of the modulator as shown in Figure~\ref{fig:1}b. Transverse magnetic modes are supported only for the 1.5-\textmu m-thin film due to the thickness of the LN film. Moreover, for the 1.5\,\textmu m thin film, the optical mode is significantly more confined within the lithium niobate, with a mode area of 6.2~\textmu m$^2$ compared to 8.89~\textmu m$^2$ for the 0.9\,\textmu m film, where the mode also tends to leak into the sapphire substrate. Thermo-optic electrodes were defined by electron-beam lithography, followed by electron-beam evaporation and gold lift-off. The electro-optic electrodes were patterned using direct laser writing, followed by another evaporation and lift-off of gold. The electro-optic electrodes are 900\,nm thick, with a cross-sectional design optimized for impedance matching to a 50\,\(\Omega\), enabling high-speed modulation. Thermo-optic electrodes were designed with a thickness of 100 nm to ensure efficient Joule heating while maintaining structural integrity and preventing electrode burnout. The facet of the device was diced, and the waveguide facets were mechanically polished to minimize in- and out-coupling losses, followed by focused ion beam (FIB) milling to further refine the facet quality, as shown in Fig.~\ref{fig:1}c. A thermo-optic phase shifter (TOPS) is included on one arm of the MZM as seen in Fig.~\ref{fig:1}d, in order to set the relative phase between the two arms and ensure that the device operates at the quadrature point.



\subsection{MZM static performance}

\begin{figure}[ht!]
    \centering
    \includegraphics[width=13.5cm]{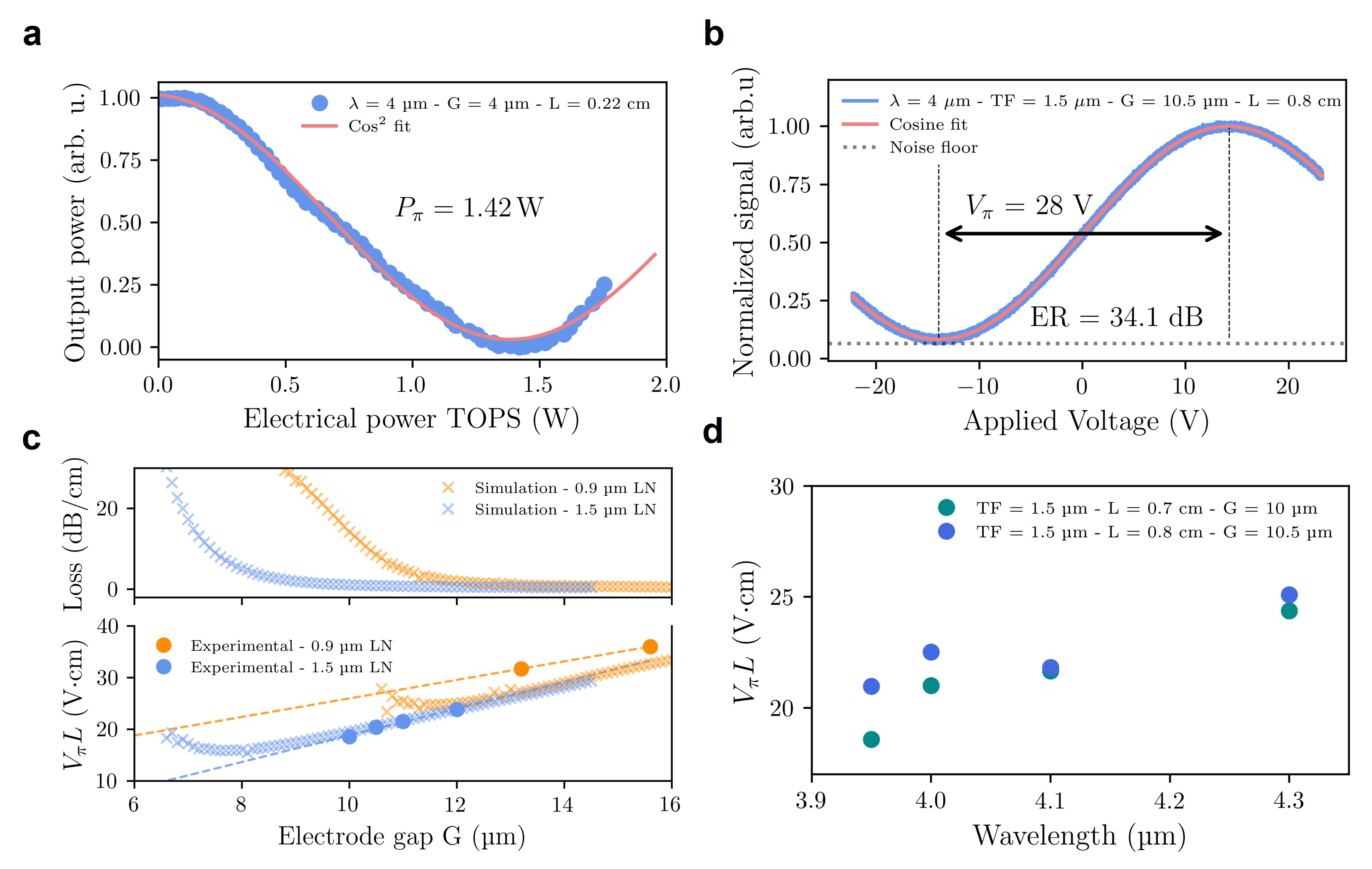}
    \caption{\textbf{Static electro-optic performance.} 
    \textbf{a,} Normalized optical transmission of the MZM as a function of the electrical power applied to the TOPS. The 4\,\textmu m transmitted power (blue) was fitted to a \( \cos^2 \) function, yielding a \( P_\pi \) of 1.43\,W for a TOPS with a length of 1.34\,mm and an electrode gap of 4\,\textmu m. \textbf{b,} Normalized optical transmission of a modulator with an 11\,\textmu m electrode gap, exhibiting a $\mathrm{V}_{\pi}$ of 28 V.cm, corresponding to a $\mathrm{V}_{\pi} \mathrm{L}$ of 22 V.cm, and an extinction ratio of 34.1 dB. \textbf{c,} Simulated and experimental results (top) for optimizing the electrode gap G between the signal and ground electrodes for lithium niobate thin films with thicknesses of 0.9\,\textmu m and 1.5\,\textmu m, operating at a wavelength of 4\,\textmu m. The bottom figure shows both the simulated and experimental $\mathrm{V}_{\pi} \mathrm{L}$ products and the simulated propagation loss as functions of the electrode gap for each platform. The experimental $\mathrm{V}_{\pi} \mathrm{L}$ values show strong agreement with the simulations, validating the accuracy of the design methodology. \textbf{d,} $\mathrm{V}_{\pi} \mathrm{L}$ evaluation of two different modulators as a function of wavelength, ranging from 3.95--4.3\,\textmu m. The value increases slightly with wavelength due to reduced confinement of the optical mode, however, the increase in $\mathrm{V}_{\pi}$ remains negligible.}
    \label{fig:3}
\end{figure}


Figure~\ref{fig:2} illustrates the experimental setup used for modulator characterization. The platform was tested at a wavelength of 4\,\textmu m using a Fabry–Pérot quantum cascade laser (QCL), configured in a custom-built external cavity to ensure single-mode emission with a tunability of approximately 150\,nm and an output power of around 20\,mW. Details on the external cavity can be found in the Supplementary information. A half-wave plate placed after the QCL rotates the polarization to enable efficient coupling into the fundamental transverse electric (TE) mode of the waveguide. The beam is coupled into a multimode mid-infrared-compatible InF$_3$ lensed fiber and then into the input waveguide. At the output, light is collected by a second InF$_3$ lensed fiber and either directly injected into an optical spectrum analyzer (OSA), power meter, or collimated and focused onto a mid-infrared photodetector using a fiber collimator and a high-numerical-aperture germanium lens. To evaluate the modulator's performance, several tests were performed. The optical path difference between the two arms is thermally tunable via the integrated TOPS, which was characterised by varying the applied electrical power and monitoring the corresponding change in the MZM output power. A half-wave power ($\mathrm{P}_{\pi}$) of 1.42\,W was extracted, as shown in Figure~\ref{fig:3}a. To extract the half-wave voltage ($\mathrm{V}_{\pi}$), the MZM was driven with a 100\,kHz triangular waveform from an arbitrary function generator and amplified using a high-voltage amplifier. The resulting optical signal was detected by a mercury cadmium telluride (MCT) detector and recorded on an oscilloscope. The measured response is presented in Figure~\ref{fig:3}b. The presented modulator exhibits a $\mathrm{V}_{\pi}L$ of 22.4\,V.cm, with an electrode length of 0.8\,cm and a gap of 10.5\,\textmu m, operating at a wavelength of 4\,\textmu m. The measured extinction ratio reaches 34.1\,dB, highlighting the high fabrication quality of the device. The modulated output power of the device reaches 350\,\textmu W, enabling straightforward characterization with a power meter, without the need for a mechanical chopper, unlike common MIR integrated platform studies. As shown in Figure~\ref{fig:3}c, simulations were performed to minimize optical propagation loss while maintaining efficient modulation performance, characterized by the half-wave voltage–length product ($\mathrm{V}_{\pi}L$), for both thin-film thicknesses. The simulated optical mode loss (see top panel of Figure~\ref{fig:3}c) increases significantly when the electrode spacing ($G$) is reduced below 12\,\textmu m and 8\,\textmu m for the 0.9\,\textmu m and 1.5\,\textmu m films, respectively, due to plasmonic absorption. The thicker 1.5\,\textmu m film provides improved optical confinement within the waveguide, whereas optical confinement in the 0.9\,\textmu m film is reduced, leading to a larger mode overlap with the sapphire substrate, which does not contribute to the modulation process. This improved confinement allows the electrodes to be placed closer to the waveguide without incurring high optical loss. Specifically, gaps of 13.2\,\textmu m and 15.6\,\textmu m were used for the 0.9\,\textmu m thin film, while gaps between 10 and 12\,\textmu m were selected for the 1.5\,\textmu m thin film. Bottom panel of Figure~\ref{fig:3}c presents both simulated and experimental results for the achieved $\mathrm{V}_{\pi}\mathrm{L}$. While the simulation shows good agreement with the experimental results for the 1.5,\textmu m devices, a slight discrepancy is observed for the 0.9,\textmu m case, likely due to a non-uniform gap length along the electrode. With the 1.5\,\textmu m film, a minimum of 18.4\,V.cm was obtained for the smallest-gap modulator, while the 0.9\,\textmu m film yielded 31.4\,V.cm. The modulator with the lowest $\mathrm{V}_{\pi}$ exhibited moderate output power of approximately 150\,\textmu W due to increased plasmonic loss caused by a slight electrode misalignment. This justified the choice to present results from the modulator with the highest optical output power instead. As shown by comparison with the simulations, conservative design choices were made to ensure low optical loss, tolerance to fabrication variations, and clear demonstration of the viability of the proposed device architecture, especially given the moderate efficiency of typical mid-infrared detectors. 
As shown in Figure~\ref{fig:3}d, we demonstrate the operational bandwidth of the MZM from 3.95\,\textmu m to 4.3\,\textmu m by evaluating the $\mathrm{V}_{\pi} \mathrm{L}$ as a function of wavelength, which remains nearly constant over this range. This corresponds to an impressive 300\,nm bandwidth. The bandwidth could potentially extend to wavelengths below 3.95\,\textmu m; however, this region could not be characterized due to the lack of available optical sources at those wavelengths in our laboratory. Finally, we measured a total fiber-to-fiber insertion loss of 14.1\,dB at 4\,\textmu m and 17.5\,dB at 4.3\,\textmu m, which includes coupling losses, waveguide propagation losses, and additional plasmonic losses due to the electrodes. Using the intercept of the cutback measurement, shown in the supplementary, we can remove the coupling contributions and evaluate the intrinsic device loss—including waveguide and plasmonic electrode losses to be at most 4\,dB at 4\,\textmu m and 5\,dB at 4.3\,\textmu m. 

\begin{figure}[ht!]
    \centering
    \includegraphics[width=13cm]{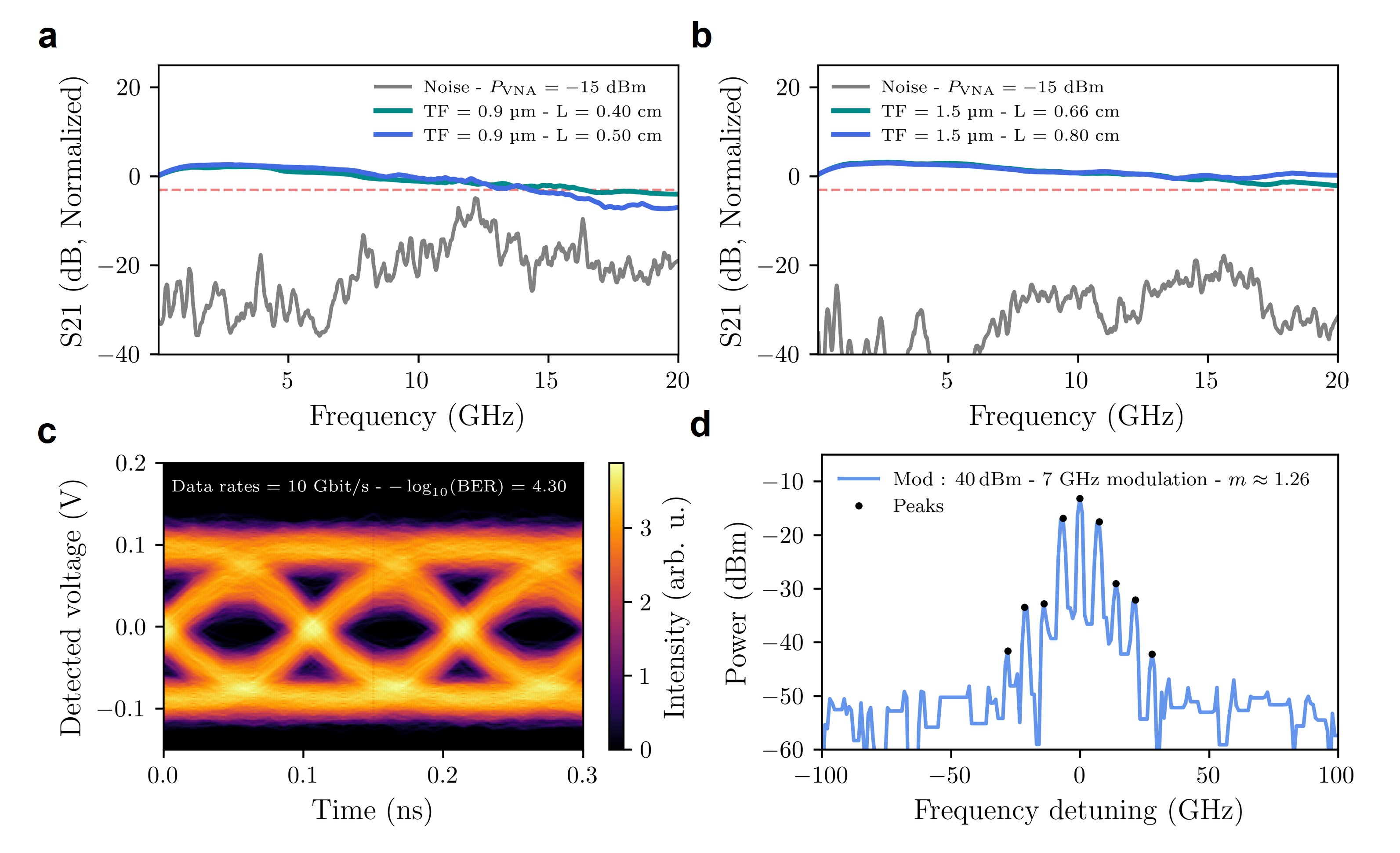}
    \caption{\textbf{High-speed electro-optic characterization of the MIR MZM.} \textbf{a,} Measured $\mathrm{S_{21}}$ electro-optic response of two MZMs with different electrode lengths using the thin-film with thickness of 0.9\,\textmu m, showing a 3 dB modulation bandwidth up to 15\,GHz. \textbf{b,} Measured $\mathrm{S_{21}}$ electro-optic response of two MIR modulators with different electrode lengths, fabricated on a 1.5\,\textmu m thin film. Both devices exhibit a flat modulation response up to 20\,GHz. Notably, the reduced $\mathrm{V}_{\pi}$ and lower coupling loss result in a significantly higher signal-to-noise ratio. \textbf{c,} An optical frequency comb was generated using 40\,dBm amplitude modulation centred around 4\,\textmu m, resulting in 8 visible sidebands spaced by 7\,GHz. The measurement was generated using an OSA with a resolution of 1.9 GHz. \textbf{d,} 10\,Gbit/s transmission using the presented modulator at 4\,\textmu m exhibited a clean eye diagram with a bit error rate below $5.10^{-5}$, corresponding to $-\log_{10}(\mathrm{BER}) = 4.3$.}
    \label{fig:4}
\end{figure}


\subsection{MZM high speed performance}

The high-speed electro-optic performance of the modulators was characterized using $\mathrm{S_{21}}$, which represents the small-signal transfer function from the electrical input to the electrical output of a photodetector monitoring the modulated optical signal. Several devices with varying film thickness and electrode lengths of L = 0.4~cm, L = 0.5~cm, L = 0.66~cm and L = 0.8~cm and gaps were evaluated. The travelling wave electrodes were externally terminated with a 50~$\Omega$ RF load for impedance matching. Electro-optic bandwidth measurements were carried out using a Vector Network Analyzer (VNA) in combination with an InAs/InAsSb type-II superlattice (T2SL) photodetector, which exhibits a bandwidth exceeding 20\,GHz. The detector demonstrates a high responsivity of over 0.8\,A/W at 4\,\textmu m when biased above 2\,V, as verified through FTIR measurements. Detailed characterization of the detector can be found in the Supplementary Material.
The VNA output was amplified using a 30\,dB RF high-power amplifier with a saturation power of approximately 20\,dBm. The 0.9\,\textmu m platform exhibited a maximum bandwidth of up to 16\,GHz, with a small dependence on electrode length, as shown in Figure~\ref{fig:4}a. The observed increase in noise around 12.5\,GHz was attributed to the onset of antenna-like behaviour, whereby the detector picks up radiation from standing waves present in the RF cables and probes. The 1.5\,\textmu m platform was subsequently engineered with optimized signal and ground pad geometries to enhance radio-frequency performance. As shown in Figure~\ref{fig:4}b, this design yields a nearly flat frequency response up to 20\,GHz, which is limited by both the VNA and the detector’s intrinsic 3\,dB bandwidth. These results suggest that the actual 3\,dB bandwidth of the modulator likely exceeds 20\,GHz. Further characterization at higher frequencies was limited by the absence of a suitable RF source and the cutoff frequency of the detector. To validate the applicability of our modulator, we conducted a transmission experiment using a two-level 10\,Gbit/s pseudo-random bit sequence of length $2^{15}$, generated by an arbitrary waveform generator (AWG) operating at a sampling rate of 20\,GS/s. To evaluate the transmission quality, we present an eye diagram along with the corresponding bit error rate as shown in the Figure~\ref{fig:4}c. The total transmitted bit sequence comprised approximately $10^{6}$ bits. The primary limitation in achieving higher bit rates was the AWG sampling rate of 20\,GS/s. Additionally, the implementation of higher-order modulation formats, such as 4-level or 8-level schemes, was constrained by the signal-to-noise ratio of our system. Using our modulator, we also successfully generated an optical frequency comb featuring ten distinct spectral lines centered around the main carrier, as measured by an optical spectrum analyzer. As shown in the Figure~\ref{fig:4}d, the comb was centered at a wavelength of approximately 4,\textmu m, with symmetric sidebands clearly resolved, indicating modulation behavior. From the relative powers of the carrier and first-order sidebands, we estimated a modulation index m of 1.26, signifying efficient energy transfer from the central mode to the sidebands. This result highlights the modulator’s capability to generate rich frequency content through high-depth amplitude modulation, demonstrating its effectiveness for mid-infrared frequency comb generation.




\section{Discussion}\label{sec12}

As demonstrated above, LNOS MZMs developed for the mid-infrared range exhibit outstanding optical modulation performance, beginning with a broad wavelength tunability demonstrated from 3.95 to 4.3\,\textmu m. In addition to this tunability, the achieved extinction ratio of 34\,dB is state-of-the-art and highly promising for spectroscopic applications. Furthermore, the optical bandwidth exceeds the current state-of-the-art in the MIR by an order of magnitude. The fabrication process demonstrates tight dimensional control, low sidewall roughness, high pattern fidelity, and reproducibility across devices. This ease of fabrication opens the door to a new generation of compact, efficient modulators or other devices for MIR spectroscopy, sensing, and free-space communication.
The observed $\mathrm{V}_{\pi} \mathrm{L}$ is higher than that of comparable modulators operating at telecom wavelengths. This behaviour arises from a fundamental limitation the Pockels effect induces a change in refractive index $\mathrm{\Delta n}$, but the resulting phase shift decreases at longer wavelengths, as described by:

\begin{equation} \label{eq:1}
    \Delta\phi = \frac{2\pi}{\lambda} \Delta n L
    \end{equation}\
    
where $\mathrm{\Delta} \phi$ is the phase shift, $\mathrm{\lambda}$ is the wavelength, and L is the interaction length.
As a result, increasing the wavelength by a factor of approximately 2.5 leads to a 3--4$\times$ increase in the intrinsic $\mathrm{V}_{\pi} \mathrm{L}$ value.
However, the measured $V_\pi$ for these devices could be reduced by further optimizing the electrode geometry and spacing, or by increasing the device length. Furthermore, as demonstrated in this study, employing a thicker LN film improves optical mode confinement, enhances modulation efficiency, and enables a reduction in $\mathrm{V}_{\pi}$, achieving a $\mathrm{V}_{\pi} \mathrm{L}$ below 20\,V$\cdot$cm. A reduced $\mathrm{V}_{\pi}$ not only enhances overall modulator performance but also enables new functionalities, such as the generation of mid-infrared ring electro-optic frequency combs. In this context, several studies have demonstrated mid-infrared comb generation through phase modulation using QCLs~\cite{hugi2012mid}, including quantum-walk QCL combs that rely purely on phase modulation \cite{heckelmann2023quantum}. Future improvements in travelling-wave electrode design could enable modulation bandwidths exceeding several tens of gigahertz, making high-speed amplitude modulation feasible. Such a modulator would pave the way for compact generation of mid-infrared pulses with durations on the order of tens of picoseconds, a capability that remains highly challenging today. Current solutions rely on bulky OPOs~\cite{pupeza2015high} or external pulse compression of QCL-based frequency combs~\cite{wang2022ultrafast, taschler2021femtosecond}. However, very recent results have shown that racetrack QCLs can generate soliton-based pulses with picosecond-scale durations~\cite{kazakov2025driven}, offering a promising alternative for compact MIR pulse generation. In parallel, our work represents a first step toward the realization of amplitude-modulated MIR frequency combs based on Kerr resonators, leveraging the strong $\chi^{(3)}$ nonlinearity of lithium niobate.


\section{Methods}\label{sec_Method}

\subsection{Device fabrication}

The chip was fabricated from a commercially available wafer with an x-cut (NGK Insulator, LTD), 900 or 1500-\textmu m LN layer and a sapphire substrate. Photonic waveguides were defined using a polymer resist with 100\,kV electron beam lithography (Raith EBPG 5200+). They are then dry etched to a depth of 1 \textmu m using Ar+ plasma reactive ion etching (Oxford Instruments PlasmaPro 100). The remaining etch mask and the redeposited material after the etching process was removed using RCA-SCA1 cleaning and buffered oxide etch. Beyond removing redeposition, the RCA-SC1 also reduces sidewall roughness as it slightly etches the LN waveguide itself. After this, the chip is annealed at 500$\,^\circ C$ under ambient conditions for 2 hours to heal any damage to the crystal caused by the dry etching process and decrease propagation losses. Thermo-optic phase shifters were then patterned by electron beam lithography to define a resist bilayer, followed by a 100 nm gold evaporation (Evatec 50) and a lift-off process. The electro-optic electrodes are then defined by direct laser writing (DWL 66+), followed by a 900 nm gold evaporation and another lift-off process. Finally, the chip was diced (Disco DAD 3221) to expose the waveguide facets, which were then mechanically polished (Allied Multiprep System 8") using diamond abrasive to minimize in- and out-coupling losses. To further improve facet quality, FIB milling (TFS Helios 5 UX) was performed as a final step.

\subsection{Design of the MZM}

The performance of passive components, including waveguides and MMIs, was estimated using an eigenmode expansion solver \textnormal{Ansys Optics}. Low-loss and single-mode operation was ensured through careful selection of waveguide width and etch depth. For the 0.9\,\textmu m film, the chosen geometry consists of a waveguide top width of \textnormal{w\textsubscript{ridge}} = 4\,\textmu m and an etch depth of \textnormal{h\textsubscript{ridge}} = 0.4\,\textmu m, while for the 1.5\,\textmu m film, the geometry is defined by \textnormal{w\textsubscript{ridge}} = 2.5\,\textmu m and \textnormal{h\textsubscript{ridge}} = 0.92\,\textmu m. Electro-optic simulations using using multiphysics and FDTD software COMSOL Multiphysics and Ansys Lumerical were used to evaluate modulation efficiency and optical field confinement, and optimize electro-optic and thermo-optic performance.

\subsection{External cavity quantum cascade laser}

To enable precise wavelength tuning around 4\,\textmu m, an external cavity configuration in the Littrow geometry was implemented. The cavity consists of a QCL coupled to a ruled reflection grating (GR2550-30035) with a dispersion of 2.86\,nm/mrad, mounted on a motorized precision rotation stage (PI RS-40). The first-order diffraction from the grating is fed back into the QCL to define the lasing wavelength, while a 50:50 beamsplitter positioned between the laser and the grating is used to extract the output power. Simultaneously, the zeroth-order reflection is used to monitor the injected signal spectrum in real time. The external cavity operates in continuous-wave mode with a tunable bandwidth of approximately 150\,nm and an output power exceeding 20\,mW, enabling fine control of the emission wavelength via adjustment of the grating angle. The output of the external cavity is then coupled into an InF$_3$ lensed fiber (Le Verre Fluoré). A schematic of the external cavity and basic characterization can be found in the Supplementary Material. For performance evaluation at 4.3\,\textmu m, a distributed feedback QCL laser with a maximum output power of approximately 40\,mW was used.

\subsection{Static Characterization}

Static characterization was performed using an arbitrary waveform generator (AFG-2125) in combination with a high-voltage amplifier (Thorlabs HVA200) to generate high-power triangular waveforms. The setup is capable of delivering a maximum electrical voltage of 200\,V. The optical output from the external cavity was injected into the system using the same experimental configuration described previously. To characterize the half-wave voltage $\mathrm{V}_{\pi}$ and the extinction ratio, an MCT detector (Vigo PVM-2TE-10.6) was used to simultaneously recover the DC and AC electrical signal from the detected modulated optical signal. The signal was recorded using an oscilloscope (TBS 2000B) and subsequently processed with a custom Python script.

\subsection{InAs/InAsSb type-II superlattice photodetector}

The uni-travelling carrier (UTC) photodetector was grown by molecular beam epitaxy (MBE) on an \textit{n}-type GaSb substrate. The layer stack consisted of a 200~nm GaSb buffer, a 300~nm InAs/AlAsSb (1.5/1.5~nm) superlattice bottom contact, a 50~nm \textit{n}-type InAsSb layer, and a 420~nm unintentionally doped InAsSb drift region. The absorption region comprised a 900~nm InAs/InAsSb (2.9/1~nm) type-II superlattice with a four-step graded doping profile. An AlAsSb/AlSb (1.22/0.61~nm) superlattice served as an electron barrier, capped by a 10~nm InAsSb top contact layer. Device mesas were defined using standard photolithography and inductively coupled plasma (ICP) dry etching. Ohmic contacts were formed by Ti/Pt/Au deposition. The device surface was then passivated with SU-8, and coplanar waveguide (CPW) pads were fabricated via air-bridging to provide a 50~\(\Omega\) impedance suitable for high-frequency measurements. At room temperature and under a $-1$~V bias, the device demonstrated a responsivity of approximately 0.6~A/W at 4.5~\textmu m. The cut-off wavelength was found to be around 5.5~\textmu m at room temperature. Temperature-dependent dark current measurements showed that the current density increased from 0.02~A/cm\(^2\) at 77~K to 3.94~A/cm\(^2\) at 300~K under a $-1$~V bias. Arrhenius analysis indicated that the dark current is dominated by diffusion processes at high temperatures, with tunneling mechanisms becoming significant below 130~K. For high-speed characterization of the modulator, devices with a diameter of 20~\textmu m were used. To facilitate integration and simplify the experimental setup, the diced device was wire-bonded to a printed circuit board (PCB) equipped with a standard SMA port, allowing for direct connection to measurement equipment. Both the on-chip and SMA-packaged photodetectors were evaluated using a Lightwave Component Analyzer (LCA) system, which consisted of a 67~GHz VNA and a 1550~nm laser. The 3-dB bandwidth of the unpackaged (on-chip) device is provided in references~\cite{huang2022high,dai2022mwir}, while the packaged device exhibited a 3-dB bandwidth of 20.3~GHz at a $-5$~V bias.

\subsection{High speed characterization}

High-speed characterization was performed using a vector network analyzer (Keysight P5004A) with a maximum bandwidth of 20\,GHz. The output port of the VNA was connected to a modulator driver amplifier (AT Microwaves AT-BBLF-0020-3022B) with a bandwidth of approximately 20\,GHz and a saturation output power of around 20\,dBm. The optical coupling setup remained the same as in the static characterization. The output light is send through a germanium fiber collimator, then focused with a germanium lens with a high numerical aperture (NA = 0.83) and a focal length of 1.873\,mm onto the $20\,\mu\mathrm{m}^2$ active area of the detector. The electrical output of the detector was then amplified using a 35\,dB low-noise amplifier (AT Microwaves AT-LNA-0043-3504Y) with a 3\,dB bandwidth of 43.5\,GHz, which was subsequently connected to the input port of the VNA. Both amplifiers and the detector were characterized independently, and their frequency responses were used to normalize the measured $\mathrm{S_{21}}$ system response, allowing recovery of the intrinsic bandwidth of the modulator. For the transmission experiment, a 20 GS/s arbitrary waveform generator (ARB Rider AWG-7204) was used in conjunction with a 30 GHz oscilloscope (Tektronix DPO77002SX). For frequency comb generation, the signal from a signal generator (Rohde \& Schwarz SGS100A) was amplified using a high-power amplifier (Fairview Microwave FMAM513), and the output was measured with a Thorlabs OSA 305 operating at a resolution of 1.9 GHz.

\backmatter

\bmhead{Supplementary information}

Further experimental data and analysis can be found in the Supplementary Information.

\bmhead{Acknowledgements}

The authors thank Granouzis Solounias Dimitrios for developing the Python program used to control the external cavity quantum cascade laser.

\bmhead{Data availability}

The data supporting the plots within this article and other findings of this study are available from the corresponding author upon reasonable request.


\noindent





\begin{appendices}

\section{Device fabrication schematic}\label{app_1}

\begin{figure}[H]
    \centering
    \includegraphics[width=13cm]{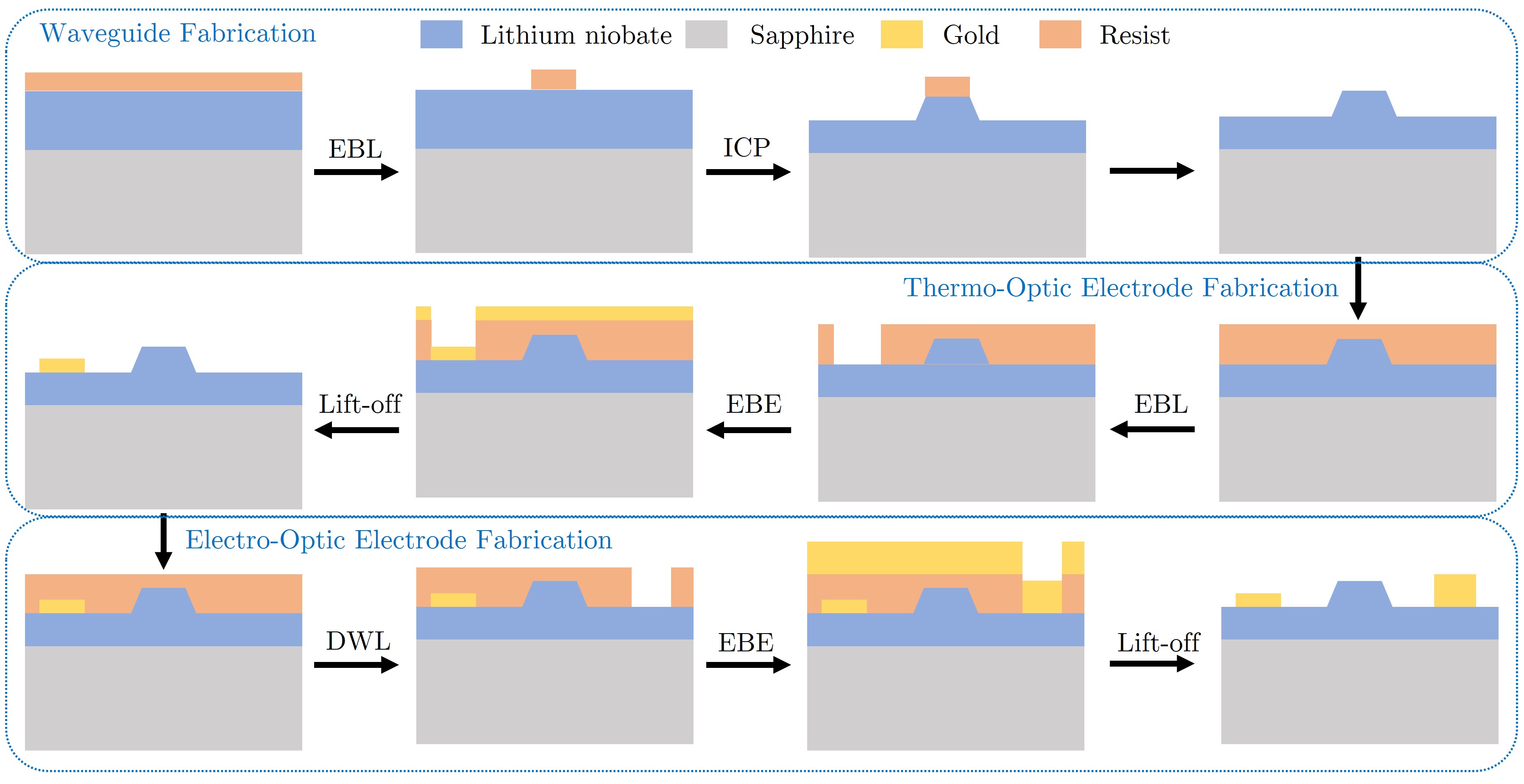}
    \caption{\textbf{Fabrication process of the TFLN on sapphire} 
    Fabrication process of the lithium niobate on sapphire. LNOS: Lithium niobate on sapphire. EBL: Electron beam lithography. ICP: Inductively Coupled Plasma. EBE: Electron beam evaporation. DWL: Direct-write laser lithography system.} 
    \label{fig:1_supp}
\end{figure}

Figure~\ref{fig:1_supp} shows a schematic representation of the fabrication process, as detailed in the Methods section, used to define photonic circuits in thin-film lithium niobate on sapphire.

\section{Quantum cascade laser external cavity}\label{app_2}

\begin{figure}[H]
    \centering
    \includegraphics[width=13cm]{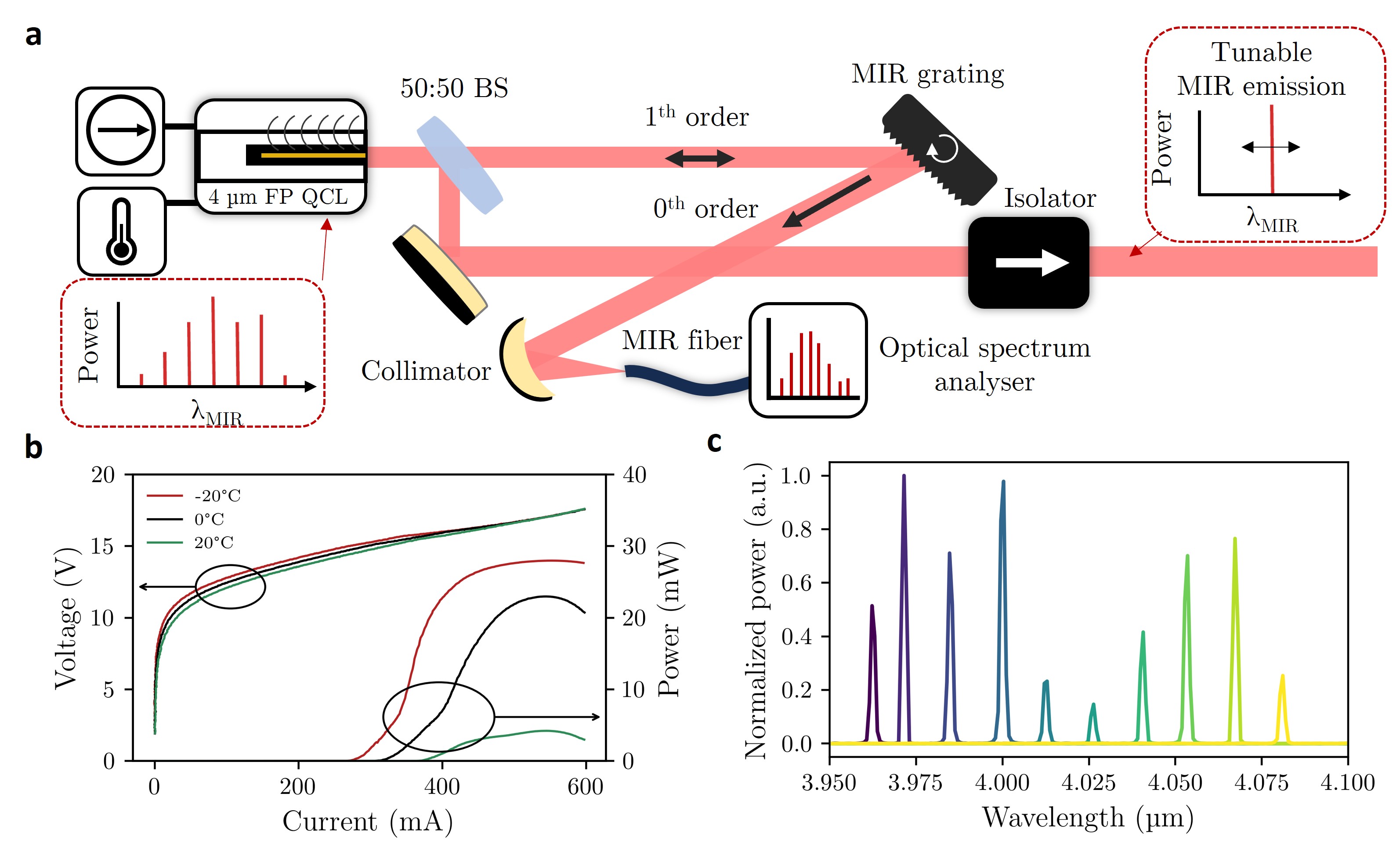}
    \caption{\textbf{Fabrication process of the TFLN on sapphire} \textbf{a,} Experimental setup of the external cavity configuration enabling the conversion from a multimode Fabry–Pérot emission at 4\,\textmu m to a tunable single-mode output. \textbf{b,} Power–voltage characteristics of the Fabry-Pérot quantum cascade laser (QCL) used in the experiment. \textbf{c,} Emission spectra of the laser, demonstrating a tunability of approximately 150 nm.} 
    \label{fig:2_supp}
\end{figure}

A schematic representation of the setup is shown in Figure~\ref{fig:2_supp}a. The 4\,\textmu m quantum cascade laser was characterized in terms of output power, intensity, and voltage, exhibiting an output power of 20\,mW at 0\,$^\circ$C with a corresponding bandwidth of approximately 150\,nm, as shown in Figures~\ref{fig:2_supp}b and~\ref{fig:2_supp}c, respectively.

\section{Cutback measurement}\label{app_3}

\begin{figure}[H]
    \centering
    \includegraphics[width=13cm]{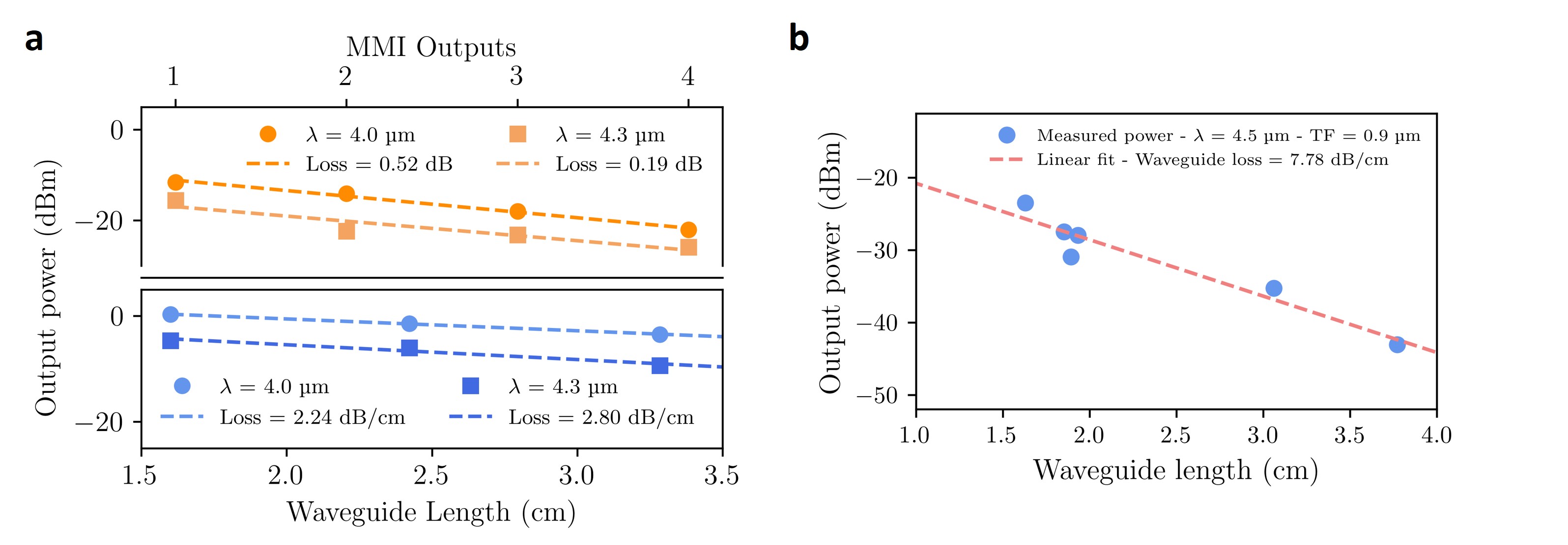}
    \caption{\textbf{Waveguide loss characterization using the cut-back measurement method.} \textbf{a,} Cut-back measurement at a wavelength of 4\,\textmu m for a waveguide with a top width of 2.5\,\textmu m and a thin film thickness of 1.5\,\textmu m. \textbf{b,} Cut-back measurement at a wavelength of 4.5\,\textmu m, showing a high loss of approximately 7.7\,dB/cm, indicating the onset of the transparency limit of lithium niobate.} 
    \label{fig:3_supp}
\end{figure}

Propagation losses were measured using waveguide lengths ranging from 1.8 to 3.8\,cm. The extracted loss is approximately 2.24\,dB/cm at 4\,\textmu m and 2.80\,dB/cm at 4.3\,\textmu m for a thin-film thickness of 1.5\,\textmu m for a single mode waveguide. These values represent an upper bound on the actual propagation loss, as bending losses were not accounted for in this study. Notably, the loss at 4.3\,\textmu m does not increase significantly with respect to the 4\,\textmu m propagation losses, supporting the feasibility of using the modulator at this wavelength. Total in- and out-coupling losses are estimated to be 8.6\,dB at 4\,\textmu m and 12.3\,dB at 4.3\,\textmu m by evaluating the linear fit of waveguide propagation losses at zero length. The performance of the MMIs was further evaluated using a four-stage cascaded MMI structure. Based on the fitting of the transmitted power data, the MMI loss is 0.52\,dB at 4\,\textmu m and 0.19\,dB at 4.3\,\textmu m, as shown in Figure~\ref{fig:3_supp}a. The latter represents the lowest reported MMI loss in this wavelength range. Interestingly, and contrary to our design expectations, the loss at 4.3\,\textmu m is lower than at 4\,\textmu m, likely due to fabrication uncertainties. Additionally, the loss at 4.5\,\textmu m was evaluated to be approximately 7.7\,dB/cm, as shown in Figure~\ref{fig:3_supp}b, which would render the use of our modulator at this wavelength unfeasible. This indicates the upper loss limit imposed by the material absorption of both the lithium niobate and sapphire substrates.

\label{app_4}

\section{High speed characterization of InAs/InAsSb type-II superlattice photodetector}

\begin{figure}[H]
    \centering
    \includegraphics[width=13cm]{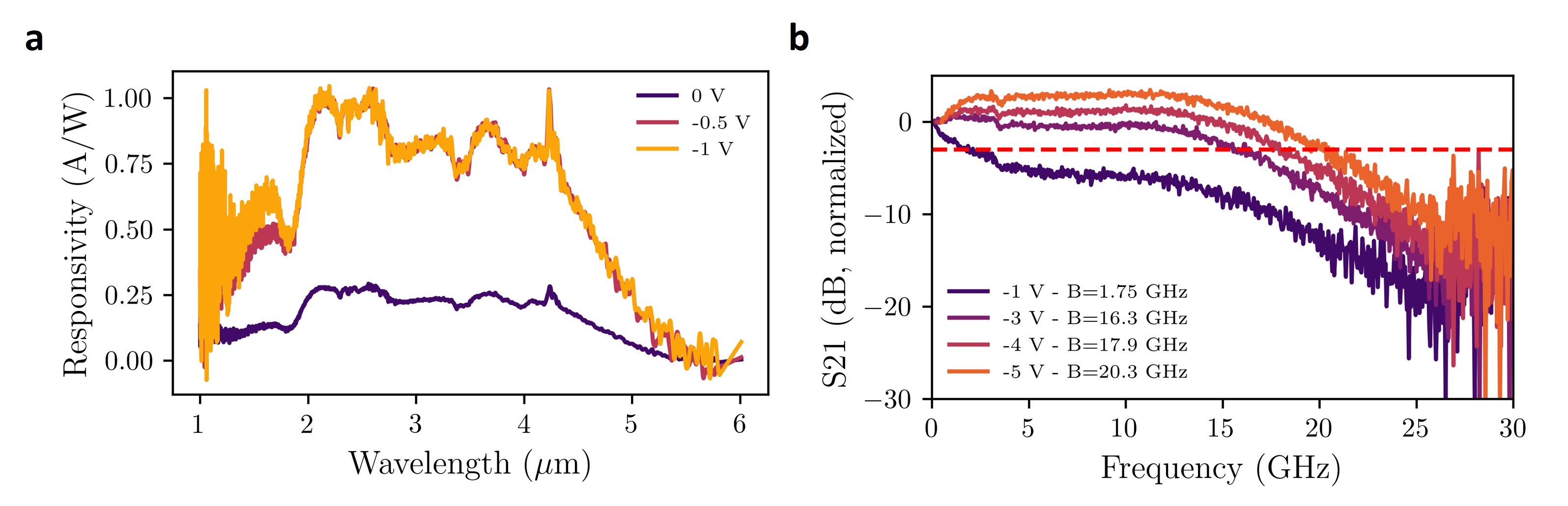}
    \caption{High-speed electro-optic characterization of the MIR PD. \textbf{a,} Responsivity of the InAs/InAsSb type-II superlattice photodetector, reaching 0.8 A/W at 4\,\textmu m, characterized via FTIR measurements under a bias voltage of 0 V, -0.5 V and -1 V. \textbf{b,} Bandwidth of the whole high-speed packaged photodetector to an SMA port, exceeding a bandwidth of 20.3 GHz under bias voltage of -5 V, characterized via an LCA system.}
    \label{fig:4_supp}
\end{figure}

For optoelectronic characterization, the responsivity of the photodetector was measured using a Fourier transform infrared spectrometer (FTIR), calibrated with a blackbody source. As shown in Figure~\ref{fig:4_supp}a, the device exhibits a room-temperature responsivity of up to 0.8~A/W at 4~µm under bias voltages of 0~V, $-0.5$~V, and $-1$~V, with the responsivity approaching saturation beyond -0.5~V. For high-speed characterization, the bandwidth of the SMA-packaged photodetector was measured using a Lightwave Component Analyzer system, comprising a 67~GHz vector network analyzer (VNA) and a 1550~nm laser. Port 1 of the VNA was used to modulate the 1550~nm laser, while Port 2 was directly connected to the photodetector’s SMA port to receive the electrical signal. The bias voltage was applied via a source meter through a bias-tee. The parameter $\mathrm{S_{21}}$ of the device was extracted to obtain the bandwidth curve shown in Figure~\ref{fig:4_supp}b. 
The results reveal a bandwidth exceeding 15~GHz at bias voltages beyond -3~V, with a maximum bandwidth of approximately 20.3~GHz at -5~V, attributed to a slight inductive peaking effect from the wire bonding.

\section{Broadband demonstration of the MZM modulator}

\begin{figure}[H]
    \centering
    \includegraphics[width=13cm]{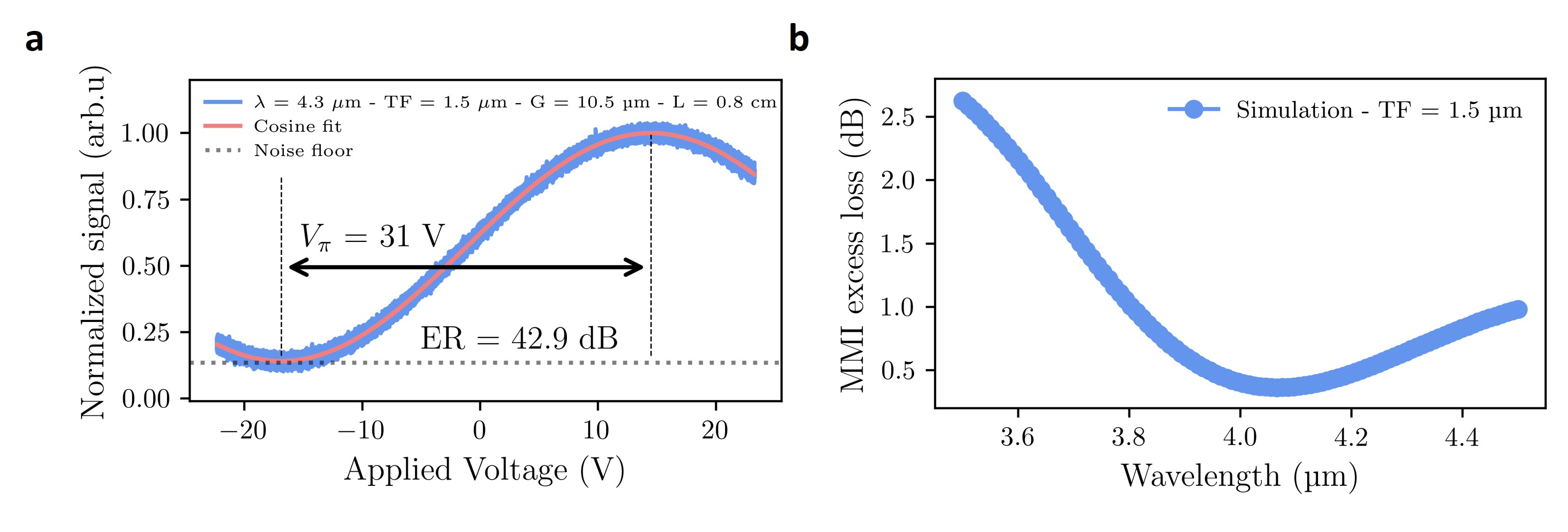}
    \caption{\textbf{a,} Normalized optical transmission of a modulator with an 11\,\textmu m electrode gap at a wavelength of 4\,\textmu m, exhibiting a $\mathrm{V}_{\pi}$ of 31\,V, corresponding to a $\mathrm{V}_{\pi} \mathrm{L}$ of 24.8\,V.cm, and an extinction ratio of 42.9\,dB. \textbf{b,} Simulated MMI excess loss as a function of wavelength from 3.5 to 4.5\,\textmu m, showing a wide bandwidth of 500\,nm for excess loss $<$ 1\,dB.}
    \label{fig:5_supp}
\end{figure}

Figure~\ref{fig:5_supp} illustrates the performance of the modulator and the broadband behavior of the MMI coupler. In Fig.\ref{fig:5_supp}.a, the normalized optical transmission of the modulator with an 11,\textmu m electrode gap is shown at a center wavelength of 4.3\,\textmu m. The modulator demonstrates efficient electro-optic modulation with a half-wave voltage of 31 V, corresponding to a voltage-length product of 24.8 V$\cdot$cm. An extinction ratio of 42.9 dB is achieved, indicating strong modulation contrast and low insertion loss. Figure.\ref{fig:5_supp}.b presents the simulated excess loss of the MMI coupler as a function of wavelength over the 3.5–4.5\,\textmu m range. The results show a wide operational bandwidth of 500 nm in which the excess loss remains below 1 dB, confirming the suitability of the MMI design for broadband mid-infrared applications.

\end{appendices}


\bibliography{sn-bibliography}

\end{document}